# WLAN Indoor Intrusion Detection Based on Deep Signal Feature Fusion and Minimized-MKMMD Transfer Learning

Mu Zhou, *Senior Member, IEEE*, Xinyue Li, Yong Wang, Aihu Ren, and Xiaolong Yang

*Abstract*—Indoor intrusion detection technology has been widely utilized in network security monitoring, smart city, entertainment games, and other fields. Most existing indoor intrusion detection methods directly exploit the Received Signal Strength (RSS) data collected by Monitor Points (MPs) and do not consider the instability of WLAN signals in the complex indoor environments. In response to this urgent problem, this paper proposes a novel WLAN indoor intrusion detection method based on deep signal feature fusion and Minimized Multiple Kernel Maximum Mean Discrepancy (Minimized-MKMMD). Firstly, the multi-branch deep convolutional neural network is used to conduct the dimensionality reduction and feature fusion of the RSS data, and the tags are obtained according to the features of the offline and online RSS fusion features that are corresponding to the silence and intrusion states, and then based on this, the source domain and target domain are constructed respectively. Secondly, the optimal transfer matrix is constructed by minimizing MKMMD. Thirdly, the transferred RSS data in the source domain is utilized for training the classifiers that are applying in getting the classification of the RSS fusion features in the target domain in the same shared subspace. Finally, the intrusion detection of the target environment is realized by iteratively updating the process above until the algorithm converges. The experimental results show that the proposed method can effectively improve the accuracy and robustness of the intrusion detection system.

## I. Introduction

WITH the increasing demand for Location-based Service (LBS), indoor intrusion detection technology plays an indispensable role in the fields of empty nesters' monitoring, valuables monitoring, and security prevention. Existing indoor intrusion detection technologies are mainly based on video surveillance [1], infrared detection [2], sensor network [3], and Wireless Local Area Network (WLAN) perception [4]. Indoor intrusion detection technology based on video surveillance has higher requirements on light conditions, which performs poorly at night or in poor light conditions. Although indoor intrusion detection technology based on infrared detection [5] is not affected by light conditions, its detection performance depends on the reliability of the Line-of-sight (LOS) paths between transmitters and receivers. Indoor intrusion detection technology based on sensor network requires deploying a large number of sensor nodes in the target environment [6], and the high deployment and maintenance cost limits the promotion and application of this technology [7]. In contrast, the method based on WLAN sensing has advantages of easy deployment, extensive coverage, and no need for special hardware equipment [8]. Therefore, it has gradually become a research hotspot of indoor intrusion detection technology.

In 2007, at the University of Maryland, Youssef et al. first proposed the concept of indoor intrusion detection based on WLAN perception [9]. In this study, there is no need for the detected target to carry any special equipment [10], and the detection process includes offline and online stages. In the offline stage, the system builds an offline database by collecting RSS data from different Access Points (APs) at each Monitor Point (MP). In the online stage, the newly collected RSS data are matched against the offline database to determine whether there is a target invasion in the target environment. Motivated from this study, in this paper, the multi-branch convolutional neural network is firstly exploited to conduct the dimension reduction and feature fusion for the RSS data extracted in the online and offline stages, and then the offline RSS fusion features with tags and online RSS with dummy tags are utilized to construct the source domain and target domain respectively. Secondly, the optimal transfer matrix based on minimizing the MKMMD between the mixed distribution of RSS fusion features in the source domain and the target domain is constructed, which is then used to transfer the RSS data in source and target domains into the same shared subspace. Thirdly, the classifiers are trained by the RSS data in the source domain, which is then used to classify the transferred RSS fusion features in the target domain to obtain the tag set for the

Manuscript received XX XX, XXXX; revised XX XX, XXXX and XX XX, XXXX; accepted XX XX, XXXX. Date of publication XX XX, XXXX; date of current version xx xx, xxxx. This work was supported in part by the Program for Changjiang Scholars and Innovative Research Team in University (IRT1299) and National Natural Science Foundation of China (61301126 and 61471077). *(corresponding author: Xinyue Li.)*
M. Zhou, Y. Wang, and Z. Tian are with the School of Communication and Information Engineering, Chongqing University of Posts and Telecommunications, Chongqing 400065, China (e-mail: zhoumu@cqupt.edu.cn; yongwang@cqupt.edu.cn; tianzs@cqupt.edu.cn).
X. Li. and A. Ren are with the School of Communication and Information Engineering, Chongqing University of Posts and Telecommunications, Chongqing 400065, China (e-mail: 443126330@qq.com; 1145962527@qq.com).

target domain. Finally, the tag set for the target domain is updated iteratively until the algorithm converges, so that the intrusion detection in the target environment is realized.

The rest of this paper is structured as follows. The second part introduces the related work on the indoor WLAN intrusion detection. The third part introduces the method proposed in this paper in detail, and the fourth part gives some relevant experimental results. Finally, the fifth part summarizes the paper and discusses the future direction.

## II. RELATED WORK

Indoor intrusion detection technology based on WLAN can detect the interference signals effectively in the target environment and protect the privacy of users. Moreover, it has a strong ability of anti-interference under the condition of Non-line-of-sight (NLOS), which has the significant practical value and broad market prospect in military and civil fields. With the rapid development of computer processing technology, data processing methods based on machine learning such as transfer learning and deep neural network have been widely used in intrusion detection systems in recent years. Among them, WLAN-based passive intrusion detection technology is a novel type of technology, which is suitable for cultural entertainment, smart home, criminal investigation [11], and other applications. In [12], the non-parametric kernel density estimation method is used to calculate the detection threshold of the offline silence data, so that when the online data features exceed the detection threshold, the system judges that there is a target intrusion. By considering the limitations of single data feature in classifiers training, the Probabilistic Neural Network (PNN) based on multiple features is exploited to improve the accuracy of intrusion detection in [13]. The authors in [14] propose a mixed learning model based on Density Peak Clustering (DPC), which improves the accuracy of the intrusion detection classification. In order to achieve dynamic targets intrusion detection, the authors in [15] propose an intrusion detection method based on Deep Neural Network (DNN), which can detect the static and dynamic intrusion targets accurately. Considering the degradation of intrusion detection performance caused by different human motion gestures, the authors in [16] utilize the fluctuation features of Channel State Information (CSI) to train a Hidden Markov Model (HMM) for human target intrusion and motion posture detection. The authors in [17] construct a new frequency-domain fingerprint by extracting the fine-grained CSI features, which is utilized to indicate the characteristic difference of indoor silence and intrusion states.

Different from the methods above, this paper proposes a new indoor intrusion detection method based on the deep signal feature fusion and minimized-MKMDD transfer learning, which can accurately determine the intrusion targets in the environment and performs steadily. In a word, the three main contributions of this paper are as follows.

• The multi-branch deep convolutional neural network is used to conduct the dimension reduction and feature fusion for RSS data to reduce the signal redundancy. Thereby, the intrusion states of the system can be described more effectively and accurately.

• The impact of signal fluctuation is minimized by calculating the MKMMD between the mixed distribution of RSS fusion features in source and target domains.

• The minimum value of MKMMD is utilized to construct the optimal transfer matrix that transfers the RSS data in source and target domains into the same shared subspace, and the detection performance is greatly improved through the iterative update.

## III. SYSTEM DESCRIPTION

Figure 1 shows the system diagram of the method proposed. Specifically, in the offline stage, the features of RSS collected at each MP are extracted, and then the feature fusion is carried out by utilizing the multi-branch deep convolutional neural network. Secondly, according to the silence and intrusion states, which are corresponding to the offline RSS fusion features, the tag set is built to construct the source domain. In the online stage, the features of online RSS are extracted to be fused by the multi-branch deep convolutional neural network. Secondly, according to the different states which are corresponding to the online RSS fusion features, the dummy tag set is built to construct the target domain. At the same time, the optimal migration matrix is constructed by minimizing the MKMMD between the mixed distributions of RSS fusion features in source and target domains, and the classifiers are trained by the data in the source domain, which is used to classify online RSS fusion features, so that the tag set for the target domain can be obtained. Based on this, through the iterative operation of the process above, when the algorithm converges, the tag set for the target domain is recognized as the intrusion detection result in the target environment.

### A. Construction of source and target domains
### 1) Feature extraction of RSS data

In the target environment, $M$ APs and $N$ MPs are deployed to obtain $p = M \times N$ signal propagation paths [18], and the sliding window function is denoted as $W_{j,t_0} = \left[ s^j_{t_0-L+1}, s^j_{t_0-L+2}, \cdots, s^j_{t_0} \right]$ where $t_0$ is the initial timestamp of the sliding window, $s^j_t$ is the RSS collected for the $j$-th path at the timestamp $t$, and $L$ is the length of the sliding window. There are eight features of the offline RSS data are extracted in each sliding window including the RSS mean, variance, maximum, minimum, range, median, RSS with the highest probability of occurrence, and probability of RSS exceeding the mean. For $p$ propagation paths, $q$ ($q = 8p$) eigenvectors can be obtained in the sliding window at the timestamp $t_0$, so that the offline RSS original feature matrix can be constructed as $\mathbf{F}_S = \left( F_1, \cdots, F_{n_s} \right)^T \in \Omega^{n_s \times q}$ where $F_s = \left( f^s_1, \cdots, f^s_q \right)$ ($s = 1, \cdots, n_s$) is the $s$-th offline RSS original eigenvector and $n_s$ is the number of the sliding windows in the offline stage.

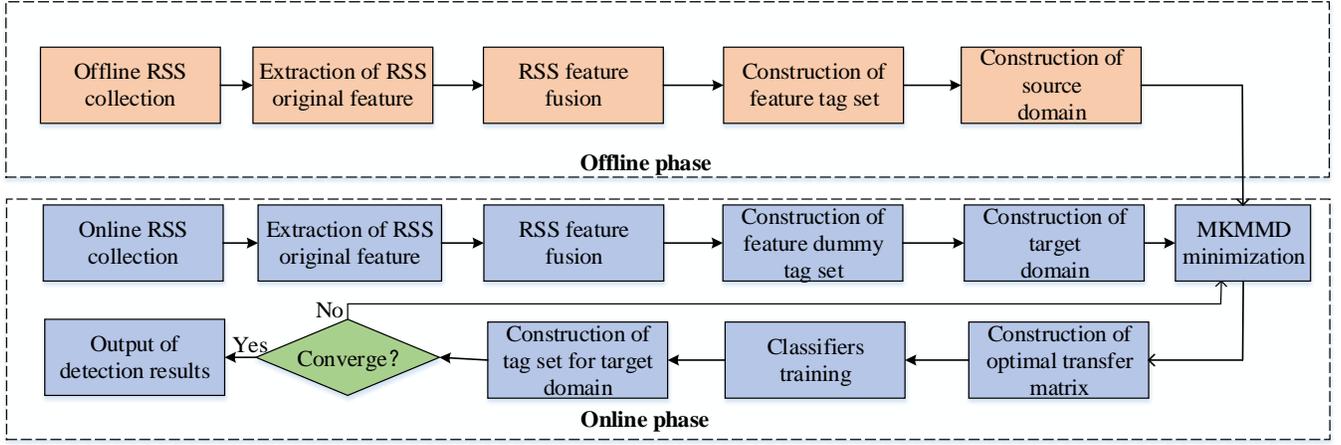

Fig. 1. System diagram.

In the online stage, the length of the sliding window is also set as $L$, which is utilized to extract the features of the online RSS data, and then the online RSS feature matrix is constructed as $\mathbf{F}_T = (\mathbf{F}'_1, \cdots, \mathbf{F}'_{n_t}) \in \Omega^{n_t \times q}$, where $\mathbf{F}'_t = (f^t_1, \cdots, f^t_q)(t = 1, \cdots, n_t)$ is the $t$-th online RSS original eigenvector and $n_t$ is the number of the sliding windows in the online stage.

*2) Deep fusion of multi-dimensional RSS features*

At present, when constructing relevant data sets for indoor intrusion detection, usually there are the problems with high dimensions of input and data redundancy [19]. Thus, in order to fully explore the useful information contained in the feature data of RSS, a multi-branch deep convolutional neural network is constructed to conduct the dimensionality reduction and feature fusion for the high-dimensional RSS original eigenvectors, which can be exploited to build the low-dimensional RSS data set.

In the same state, the original eigenvector $\mathbf{F}_s = (f^s_1, \cdots, f^s_q)$ that is collected by MPs is normalized as in (1), and the synchronization of eigenvectors is guaranteed meanwhile.

$$s'_n = \frac{s_n - \min\{\mathbf{F}_s\}}{\max\{\mathbf{F}_s\} - \min\{\mathbf{F}_s\}} \quad (1)$$

where $s'_n$ is the normalized parameter of the original eigenvector $\mathbf{F}_s$. Thus, the normalized feature parameter of RSS eigenvectors, $\mathbf{F}_s$ can be obtained by traversing all original eigenvectors, denoted as $\mathbf{F}'_s = (s'_1, s'_2, \cdots, s'_q)$.

In order to extract the useful features of RSS data fully, we input the normalized feature parameters into the multi-branch deep convolutional neural network and conduct deep feature fusion. As shown in Figure 2, eight single parameter networks are fused into an end-to-end convolutional neural network [20], which independently extracts the useful information of each kind of data through multiple branches. This network can be divided into two parts. The front part selects the first six layers of the single parameter network as the branch network for information extraction, which is utilized to obtain the useful features of the original RSS data, while the latter part is to reduce the dimensions of RSS data and fuse useful features.

*3) Construction of source and target domains*

Based on the data processing above, we can obtain the fusion feature matrix of RSS data in the offline stage, which is denoted as $\mathbf{X}_S = (\mathbf{x}_1, \cdots, \mathbf{x}_{n_s})^T$. When there are $K$ states in the target environment including one silence state and $K - 1$ invasion states, the tag $y_s = k$ means that the $s$-th offline RSS fusion eigenvector is corresponding to the $k(k = 1, \cdots, K)$-th state, and then the tag set $\mathbf{y}_S = (y_1, \cdots, y_{n_s})^T$ for the offline RSS fusion feature is obtained to build the source domain $\mathcal{D}_S = \{(\mathbf{x}_1, y_1), \cdots, (\mathbf{x}_{n_s}, y_{n_s})\}$. In the online stage, the fusion feature matrix $\mathbf{X}_T = (\mathbf{x}'_1, \cdots, \mathbf{x}'_{n_t})^T$ of RSS data in the online stage can be obtained similarly. In order to calculate the MKMMD between the mixed distributions of RSS fusion features in the source and target domains, $\tilde{\mathbf{y}}_T = (y'_1, \cdots, y'_{n_t})^T$ is set as the dummy tags of RSS features to construct the target domain as $\mathcal{D}_T = \{(\mathbf{x}'_1, y'_1), \cdots, (\mathbf{x}'_{n_t}, y'_{n_t})\}$ [21].

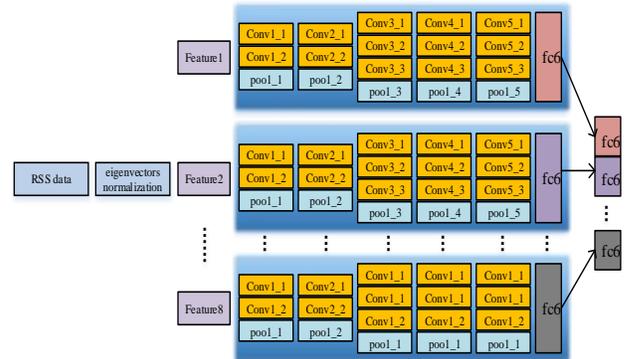

Fig. 2. Structure diagram of multi-branch deep convolutional neural network.

*B. Construction of optimal transfer matrix*

In the indoor intrusion detection method based on database construction, the classifiers obtained by learning the RSS fusion feature in the source domain can classify the RSS features in the target domain to judge whether there is an intrusion in the target environment. However, the time-varying WLAN signal in a complex indoor environment may cause a significant difference in the distribution of RSS features between the source and target domains even at the same location [22], which results in the poor robustness of the WLAN indoor intrusion detection system. Thus, choosing an appropriate measurement method to reduce the difference of RSS features distribution between the source and target domains at the same location is the key to ensure the performance of the WLAN indoor intrusion detection system. For this purpose, the authors in [23] propose to utilize the parametric methods such as Kullback-Leibler divergence to measure the difference between any two distributions, which needs to estimate the density of the data, but the accuracy of the density estimation is difficult to be guaranteed well in the actual environment. In response to this compelling problem, the authors in [24] propose to use non-parametric methods such as Maximum Mean Discrepancy (MMD) to measure the difference between different distributions. Compared with the parametric method, the non-parametric method measures the difference between the corresponding distributions by calculating the mean distance of the data from different distributions in the Reproducing Kernel Hilbert Space (RKHS) [25]. Based on this, considering the advantage of the MKMMD in reducing the second kind of error probability, which is defined as the probability of erroneously accepting the original hypothesis that the data follows the same distribution when the data distribution is different, the authors in [26] propose to minimize the MKMMD of the RSS feature distributions between source and target domains with the purpose of enhancing the robustness of WLAN indoor intrusion detection systems. The specific process of this method is described as follows.

Let $P(\mathbf{X}_S)$ and $P(\mathbf{X}_T)$ be the marginal distributions of RSS fusion features in source and target domains respectively, and their MMD can be expressed as

$$D(P(\mathbf{X}_S), P(\mathbf{X}_T)) = \left\| \mathrm{E}_{P(\mathbf{X}_S)}(\phi(\mathbf{X}_S)) - \mathrm{E}_{P(\mathbf{X}_T)}(\phi(\mathbf{X}_T)) \right\|_{\mathcal{H}}^2 \quad (2)$$

where $\mathrm{E}_{P(\mathbf{X}_S)}(\cdot)$ and $\mathrm{E}_{P(\mathbf{X}_T)}(\cdot)$ represent the expectation operations when the marginal distributions of $\mathbf{X}_S$ and $\mathbf{X}_T$ are $P(\mathbf{X}_S)$ and $P(\mathbf{X}_T)$ respectively, $\|\cdot\|_{\mathcal{H}}^2$ represents the 2-norm operation in the RKHS, and $\phi$ represents the mapping function that maps the fusion feature matrix of RSS into RKHS [27]. Since it is difficult to calculate the population mean of the whole RSS in source and target domains (i.e., expectation operation), in this paper the equation above is approximated by calculating the mean of RSS samples in source and target domains in (3).

$$D(P(\mathbf{X}_S), P(\mathbf{X}_T)) \approx \left\| \frac{1}{n_s} \sum_{i=1}^{n_s} \phi(\mathbf{x}_i) - \frac{1}{n_t} \sum_{i=1}^{n_t} \phi(\mathbf{x}'_i) \right\|_{\mathcal{H}}^2 \quad (3)$$

Although the MMD between the marginal distribution of RSS fusion features in the source domain and the marginal distribution of RSS features in the target domain can reduce the overall difference of RSS features in both source and target domains, the correlation of RSS features between different classes [28] is ignored. Thus, this paper considers both the MMD between the conditional distribution of RSS features in source and target domains. For this reason, the RSS fusion features with the same tags in source and target domains are classified into one category, and $K$ types of RSS features can be obtained respectively. By considering the calculation complexity of the posterior probability $P(\mathbf{y}_S|\mathbf{X}_S)$ and $P(\tilde{\mathbf{y}}_T|\mathbf{X}_T)$, $P(\mathbf{X}_S|\mathbf{y}_S)$ and $P(\mathbf{X}_T|\tilde{\mathbf{y}}_T)$ are taken as the conditional distributions of RSS fusion features in source and target domains respectively, and their MMD is approximated as

$$D(P(\mathbf{X}_S|\mathbf{y}_S), P(\mathbf{X}_T|\tilde{\mathbf{y}}_T)) \approx \sum_{k=1}^{K} \left\| \frac{1}{n_s^k} \sum_{i=1}^{n_s^k} \phi(\mathbf{x}_i) - \frac{1}{n_t^k} \sum_{i=1}^{n_t^k} \phi(\mathbf{x}'_i) \right\|_{\mathcal{H}}^2 \quad (4)$$

where $\mathbf{x}_i \in \mathbf{x}_s^k$, $\mathbf{x}'_i \in \mathbf{x}_t^k$, $\mathbf{x}_s^k$ and $\mathbf{x}_t^k$ are the RSS eigenvectors with the category $k$ in source and target domains respectively, and $n_s^k$ and $n_t^k$ are the number of RSS eigenvectors with the category $k$ in source and target domains respectively.

According to (3) and (4), the MMD between the mixed distributions of RSS fusion features in source and target domains can be constructed as

$$D(\mathbf{X}_S, \mathbf{X}_T) = D(P(\mathbf{X}_S), P(\mathbf{X}_S)) + D(P(\mathbf{X}_S|\mathbf{y}_S), P(\mathbf{X}_T|\tilde{\mathbf{y}}_T))$$
$$\approx \left\| \frac{1}{n_s} \sum_{i=1}^{n_s} \phi(\mathbf{x}_i) - \frac{1}{n_t} \sum_{i=1}^{n_t} \phi(\mathbf{x}'_i) \right\|_{\mathcal{H}}^2 + \sum_{k=1}^{K} \left\| \frac{1}{n_s^k} \sum_{i=1}^{n_s^k} \phi(\mathbf{x}_i) - \frac{1}{n_t^k} \sum_{i=1}^{n_t^k} \phi(\mathbf{x}'_i) \right\|_{\mathcal{H}}^2 \quad (5)$$

In this case, the optimization problem of the mapping function $\phi$ which minimizes the value of (4) can be described as

$$\min_\phi D(\mathbf{X}_S, \mathbf{X}_T) = \min_\phi \left( D(P(\mathbf{X}_S), P(\mathbf{X}_S)) + D(P(\mathbf{X}_S|\mathbf{y}_S), P(\mathbf{X}_T|\tilde{\mathbf{y}}_T)) \right)$$
$$\approx \min_\phi \left\| \frac{1}{n_s} \sum_{i=1}^{n_s} \phi(\mathbf{x}_i) - \frac{1}{n_t} \sum_{i=1}^{n_t} \phi(\mathbf{x}'_i) \right\|_{\mathcal{H}}^2$$
$$+ \sum_{k=1}^{K} \left\| \frac{1}{n_s^k} \sum_{i=1}^{n_s^k} \phi(\mathbf{x}_i) - \frac{1}{n_t^k} \sum_{i=1}^{n_t^k} \phi(\mathbf{x}'_i) \right\|_{\mathcal{H}}^2$$
$$= \min_{\mathbf{K}} \left( \mathrm{tr}(\tilde{\mathbf{K}}\mathbf{L}_0) + \sum_{k=1}^{K} \mathrm{tr}(\tilde{\mathbf{K}}\mathbf{L}_k) \right) = \min_{\mathbf{K}} \sum_{k=0}^{K} \mathrm{tr}(\tilde{\mathbf{K}}\mathbf{L}_k)$$
(6)

where the notation "tr" means the trace operation of the matrix, the element on the $i$-th row and $j$-th column of $\tilde{\mathbf{K}} \in \Omega^{(n_s+n_t) \times (n_s+n_t)}$ is $(\tilde{\mathbf{K}})_{ij} = \phi(\mathbf{x}_i)^\mathrm{T} \phi(\mathbf{x}_j)$ ($1 \leq i \leq n_s + n_t; 1 \leq j \leq n_s + n_t$; $\mathbf{x}_i, \mathbf{x}_j \in \mathbf{X}_S \cup \mathbf{X}_T$) the element on the $i$-th row and $j$-th column of $\mathbf{L}_0$ is $(\mathbf{L}_0)_{ij} = \begin{cases} 1/(n_s)^2, & \text{if } \mathbf{x}_i, \mathbf{x}_j \in \mathbf{X}_S \\ 1/(n_t)^2, & \text{if } \mathbf{x}_i, \mathbf{x}_j \in \mathbf{X}_T \\ -1/(n_s n_t), & otherwise \end{cases}$, and the element on the $i$-th row and $j$-th column of $\mathbf{L}_k$ ($1 \leq k \leq K$) is

$$(\mathbf{L}_k)_{ij} = \begin{cases} 1/(n_s^k)^2, & \text{if } \mathbf{x}_i, \mathbf{x}_j \in \mathbf{x}_s^k \\ 1/(n_t^k)^2, & \text{if } \mathbf{x}_i, \mathbf{x}_j \in \mathbf{x}_t^k \\ -1/(n_s^k n_t^k), & \text{if } \mathbf{x}_i \in \mathbf{x}_s^k \cup \mathbf{x}_j \in \mathbf{x}_t^k \text{ or } \mathbf{x}_i \in \mathbf{x}_t^k \cup \mathbf{x}_j \in \mathbf{x}_s^k \\ 0, & otherwise \end{cases}.$$

The solution of (5) is usually obtained by adopting the Semi-definite Program (SDP) method with significant computational overhead [29]. In order to reduce the overhead, the kernel matrix construction method is exploited to solve the formula above. To achieve this goal, we define the kernel matrix $\mathbf{K} \in \Omega^{(n_s+n_t)\times(n_s+n_t)}$ as

$$\mathbf{K} = \begin{bmatrix} \mathbf{K}_{s,s} & \mathbf{K}_{s,t} \\ \mathbf{K}_{t,s} & \mathbf{K}_{t,t} \end{bmatrix} \quad (7)$$

where $\mathbf{K}_{s,s}$, $\mathbf{K}_{t,t}$, and $\mathbf{K}_{s,t} (=\mathbf{K}_{t,s}^T)$ are the Gram matrices in the source domain [30], target domain, and mixed domain (including source and target domains) respectively, and the elements on the $i$-th row and $j$-th column of $(\mathbf{K}_{s,s})_{i,j} = \mathbf{x}_i \mathbf{x}_j^T (\mathbf{x}_i, \mathbf{x}_j \in \mathbf{X}_S)$, $(\mathbf{K}_{t,t})_{i,j} = \mathbf{x}_i \mathbf{x}_j^T (\mathbf{x}_i, \mathbf{x}_j \in \mathbf{X}_T)$, and $(\mathbf{K}_{s,t})_{i,j} = \mathbf{x}_i \mathbf{x}_j^T (\mathbf{x}_i \in \mathbf{X}_S, \mathbf{x}_j \in \mathbf{X}_T)$ are $\mathbf{K}_{s,s}$, $\mathbf{K}_{t,t}$, and $\mathbf{K}_{s,t} (=\mathbf{K}_{t,s}^T)$ respectively. By utilizing the empirical kernel mapping method [31], $\mathbf{K}$ is decomposed into

$$\mathbf{K} = (\mathbf{K}\mathbf{K}^{-1/2})(\mathbf{K}^{-1/2}\mathbf{K}) \quad (8)$$

Letting $\mathbf{W} = \mathbf{K}^{-1/2}\tilde{\mathbf{W}}$, where $\tilde{\mathbf{W}} \in \Omega^{(n_s+n_t)\times q} (q \le p)$ is the transformation matrix, $\tilde{\mathbf{K}}$ can be expressed as

$$\tilde{\mathbf{K}} = (\mathbf{K}\mathbf{K}^{-1/2}\tilde{\mathbf{W}})(\tilde{\mathbf{W}}^T\mathbf{K}^{-1/2}\mathbf{K}) = \mathbf{K}\mathbf{W}\mathbf{W}^T\mathbf{K} \quad (9)$$

Thus, the MMD between the mixed distributions of RSS features in source and target domains can be rewritten as

$$\begin{aligned} D(\mathbf{X}_S, \mathbf{X}_T) &\approx \sum_{k=0}^{K} \mathrm{tr}(\tilde{\mathbf{K}}\mathbf{L}_k) \\ &= \sum_{k=0}^{K} \mathrm{tr}((\mathbf{K}\mathbf{W}\mathbf{W}^T\mathbf{K})\mathbf{L}_k) = \sum_{k=0}^{K} \mathrm{tr}(\mathbf{W}^T\mathbf{K}\mathbf{L}_k\mathbf{K}\mathbf{W}) \end{aligned} \quad (10)$$

Besides, by considering the multi-core architecture [32], the subspace mapping of multiple core functions is combined to improve the mapping ability of different features in each subspace [33], so that the RSS features in the combination space can be expressed more accurately and reasonably. Based on this, the positive definite kernel function is defined as $f = \sum_{g=1}^{G} \alpha_g f_g$, where $f_g$ is the $g (g=1,\cdots,G)$-th kernel function, $\alpha_g (\ge 0)$ is the $g$-th constant coefficient, $\sum_{g=1}^{G} \alpha_g = 1$, and $G$ is the number of kernel functions. Here, we let $\mathbf{K}_g = \begin{bmatrix} \mathbf{K}_{s,s}^g & \mathbf{K}_{s,t}^g \\ \mathbf{K}_{t,s}^g & \mathbf{K}_{t,t}^g \end{bmatrix}$, where the element on the $i$-th row and $j$-th column of $\mathbf{K}_{s,s}^g$, $\mathbf{K}_{t,t}^g$, and $\mathbf{K}_{s,t}^g$ are $(\mathbf{K}_{s,s}^g)_{ij} = f_g(\mathbf{x}_i, \mathbf{x}_j)(\mathbf{x}_i, \mathbf{x}_j \in \mathbf{X}_S)$, $(\mathbf{K}_{t,t}^g)_{ij} = f_g(\mathbf{x}_i, \mathbf{x}_j \in \mathbf{X}_T)$ and $(\mathbf{K}_{s,t}^g)_{ij} = f_g(\mathbf{x}_i, \mathbf{x}_j)(\mathbf{x}_i \in \mathbf{X}_S, \mathbf{x}_j \in \mathbf{X}_T)(=\mathbf{K}_{t,s}^{g\,T})$ respectively.

Thus, the minimization objective function of the MKMMD between the mixed distributions of RSS features in source and target domains can be constructed as

$$\min_{\mathbf{W}} \sum_{k=0}^{K} \mathrm{tr}\left(\mathbf{W}^T \left(\sum_{g=1}^{G} \alpha_g \mathbf{K}_g\right) \mathbf{L}_k \left(\sum_{g=1}^{G} \alpha_g \mathbf{K}_g\right) \mathbf{W}\right) \quad (11)$$

Since $\sum_{g=1}^{G} \alpha_g \mathbf{K}_g \mathbf{W}$ is the fusion feature of RSS after the migration, it can be obtained that the optimization problem with respect to the mapping function $\phi$ in (6) is equivalent to the one with respect to the migration matrix $\mathbf{W}$ in (11). However, due to the fact that the variance of the actual collected signal RSS is often larger than that of the noise, only minimizing the MKMMD may preserve many noise components. Thus, in order to minimize the MKMMD while preserving the distribution characteristics of RSS data [16], we rewrite (11) as

$$\begin{cases} \min_{\mathbf{W}} \sum_{k=0}^{K} \mathrm{tr}\left(\mathbf{W}^T \left(\sum_{g=1}^{G} \alpha_g \mathbf{K}_g\right) \mathbf{L}_k \left(\sum_{g=1}^{G} \alpha_g \mathbf{K}_g\right) \mathbf{W}\right) + \lambda \mathrm{tr}(\mathbf{W}^T\mathbf{W}) \\ \mathrm{s.t.} \ \mathbf{W}^T \left(\sum_{g=1}^{G} \alpha_g \mathbf{K}_g\right) \mathbf{H} \left(\sum_{g=1}^{G} \alpha_g \mathbf{K}_g\right) \mathbf{W} = \mathbf{I} \end{cases} \quad (12)$$

where $\mathbf{I}$ is the unit matrix, $\mathbf{H} = \mathbf{I} - 1/(n_s+n_t)\mathbf{e}\mathbf{e}^T$ is the central matrix, $\mathbf{e}$ is the column vector with all the elements equaling to 1, $\lambda (>0)$ is the tradeoff coefficient, $\mathrm{tr}(\mathbf{W}^T\mathbf{W})$ is the regular term, and $\mathbf{W}^T \left(\sum_{g=1}^{G} \alpha_g \mathbf{K}_g\right) \mathbf{H} \left(\sum_{g=1}^{G} \alpha_g \mathbf{K}_g\right) \mathbf{W}$ is the fusion feature divergence matrix of RSS after the migration. In this circumstance, the Lagrange multiplier method can be utilized to obtain

$$\begin{aligned} L(\mathbf{W}) &= \mathbf{W}^T \left(\sum_{g=1}^{G} \alpha_g \mathbf{K}_g\right) \sum_{k=0}^{K} \mathbf{L}_k \left(\sum_{g=1}^{G} \alpha_g \mathbf{K}_g\right) \mathbf{W} \\ &+ \lambda \mathbf{W}^T \mathbf{W} - \left(\mathbf{I} - \mathbf{W}^T \left(\sum_{g=1}^{G} \alpha_g \mathbf{K}_g\right) \mathbf{H} \left(\sum_{g=1}^{G} \alpha_g \mathbf{K}_g\right) \mathbf{W}\right) \mathbf{Z} \end{aligned} \quad (13)$$

where $\mathbf{Z}$ is the diagonal matrix formed by Lagrangian multipliers [34]. The partial derivative of $L(\mathbf{W})$ with respect to $\mathbf{W}$ is obtained as

$$\begin{aligned} \frac{\partial L(\mathbf{W})}{\partial \mathbf{W}} &= 2\left(\mathbf{W}^T \left(\sum_{g=1}^{G} \alpha_g \mathbf{K}_g\right) \sum_{k=0}^{K} \mathbf{L}_k \left(\sum_{g=1}^{G} \alpha_g \mathbf{K}_g\right) + \lambda \mathbf{I}\right) \mathbf{W} \\ &- 2\left(\sum_{g=1}^{G} \alpha_g \mathbf{K}_g\right) \mathbf{H} \left(\sum_{g=1}^{G} \alpha_g \mathbf{K}_g\right) \mathbf{W}\mathbf{Z} \end{aligned} \quad (14)$$

Letting (14) equal to zero, we obtain

$$\begin{aligned} &\left(\left(\sum_{g=1}^{G} \alpha_g \mathbf{K}_g\right) \sum_{k=0}^{K} \mathbf{L}_k \left(\sum_{g=1}^{G} \alpha_g \mathbf{K}_g\right) + \lambda \mathbf{I}\right) \mathbf{W} \\ &= \left(\sum_{g=1}^{G} \alpha_g \mathbf{K}_g\right) \mathbf{H} \left(\sum_{g=1}^{G} \alpha_g \mathbf{K}_g\right) \mathbf{W}\mathbf{Z} \end{aligned} \quad (15)$$

By multiplying $\mathbf{W}^T$ on both sides of (15), one has

$$\mathbf{W}^T \left( \left( \sum_{g=1}^{G} \alpha_g \mathbf{K}_g \right) \sum_{k=0}^{K} \mathbf{L}_k \left( \sum_{g=1}^{G} \alpha_g \mathbf{K}_g \right) + \lambda \mathbf{I} \right) \mathbf{W} \quad (16)$$

$$= \mathbf{W}^T \left( \sum_{g=1}^{G} \alpha_g \mathbf{K}_g \right) \mathbf{H} \left( \sum_{g=1}^{G} \alpha_g \mathbf{K}_g \right) \mathbf{W} \mathbf{Z} = \mathbf{Z} \mathbf{I}$$

According to (12) and (16), $\operatorname{tr}\left( \mathbf{W}^T \left( \left( \sum_{g=1}^{G} \alpha_g \mathbf{K}_g \right) \sum_{k=0}^{K} \mathbf{L}_k + \lambda \mathbf{I} \right) \right)$ $\mathbf{W}$ can be minimized by the elements in the minimization matrix $\mathbf{Z}$, and thereby the optimal migration matrix $\mathbf{W}$ can be composed of the eigenvectors corresponding to the $q$ nonzero minimum generalized eigenvalues of $\left( \sum_{g=1}^{G} \alpha_g \mathbf{K}_g \right) \sum_{k=0}^{K} \mathbf{L}_k \left( \sum_{g=1}^{G} \alpha_g \mathbf{K}_g \right) + \lambda \mathbf{I}$ with respect to $\left( \sum_{g=1}^{G} \alpha_g \mathbf{K}_g \right) \mathbf{H} \left( \sum_{g=1}^{G} \alpha_g \mathbf{K}_g \right)$. The optimal migration matrix is exploited to minimize the MKMMD between the mixed distributions of RSS fusion features in source and target domains, which can make the RSS fusion features in source and target domains migrated into the same subspace have as little difference as possible. At the same time, in this subspace, the RSS fusion features in the source domain after the migration and its corresponding tags are utilized to train the classifiers, and the trained classifiers are applied in classifying the RSS fusion features in the target domain to obtain the tag set for test samples.

*C. Acquisition of tag set for target domain*

In order to obtain a reliable tag set for the target domain and improve the accuracy of indoor intrusion detection in the WLAN, we migrate the RSS fusion features in source and target domains into the same subspace by iteratively updating the tag set in the target domain [35]. To be specific, firstly, source and target domains are constructed respectively by utilizing the fusion features of offline RSS with tags and online RSS with dummy tags. Secondly, the MKMMD between the mixed distributions of RSS fusion features in source and target domains is calculated, and the optimal migration matrix with the minimum MKMMD is constructed. Thirdly, the RSS fusion features in source and target domains are transferred into the same subspace through the optimal migration matrix [36]. At the same time, the RSS fusion features in the source domain after the migration and their corresponding tags are used to train classifiers, and the trained classifiers are used to classify the RSS fusion features in the target domain to obtain the tag set for test samples. Finally, the process above is repeated until the algorithm converges, so that the tag set for the test samples in the final target domain is obtained, which implements the result of intrusion detection in the target environment. The pseudo-code of the proposed algorithm above is shown in Algorithm 1.

---

**Algorithm 1** Iterative updating of the tag set for the target domain

---

**Input:** $\mathbf{X}_S$ (Offline RSS fusion feature matrix), $\mathbf{y}_S$ (Tag set for offline RSS fusion feature), $\mathbf{X}_T$ (Online RSS fusion feature matrix), $\tilde{\mathbf{y}}_T$ (Dummy tags for online RSS fusion features), $p$ (Number of RSS features), $q$ (Dimensions of RSS features after the migration), $\lambda$ (Tradeoff coefficient), $G$ (Number of kernel functions), $f_g$ (g-th kernel function), and $N$ (g-th kernel function)

**Output:** $\mathbf{y}_T$ (Tag set for the target domain)

1: Calculate $\sum_{g=1}^{G} \alpha_g \mathbf{K}_g$ (with the complexity of $O\left( G \left( n_s + n_t \right)^2 \right)$) and $\mathbf{L}_0$ (with the complexity of $O\left( Nq \left( n_s + n_t \right)^2 \right)$)

2: **repeat**

3: Merge the RSS fusion features with the same tags in source and target domains into the same class, and then obtain $K$ classes of RSS features

4: Calculate $\sum_{k=1}^{K} \mathbf{L}_k$ (with the complexity of $O\left( NK^2 \left( n_s + n_t \right)^2 \right)$)

5: Construct the optimal migration matrix $\mathbf{W}$ which is composed of the eigenvectors corresponding to the $q$ nonzero minimum generalized eigenvalues of $\left( \sum_{g=1}^{G} \alpha_g \mathbf{K}_g \right) \sum_{k=0}^{K} \mathbf{L}_k \left( \sum_{g=1}^{G} \alpha_g \mathbf{K}_g \right) + \lambda \mathbf{I}$ with respect to $\left( \sum_{g=1}^{G} \alpha_g \mathbf{K}_g \right) \mathbf{H} \left( \sum_{g=1}^{G} \alpha_g \mathbf{K}_g \right)$ (with the complexity of $O\left( Nq \left( n_s + n_t \right)^2 \right)$).

6: Transfer the RSS fusion features in both source and target domains into the same subspace, and then the transferred RSS fusion features $\sum_{g=1}^{G} \alpha_g \mathbf{K}_g \mathbf{W}$ are obtained

7: Utilize the RSS fusion features in the source domain after the migration and the corresponding tags for training classifiers, and then the trained classifiers are applied in classifying the RSS fusion features in the target domain to obtain the tag set for test samples

8: **until** the algorithm converges

9: Output the tag set for the final target domain $\mathbf{y}_T$

## IV. EXPERIMENTAL RESULTS

### A. Experimental layout

In the experimental environment shown in figure 3, two corridors, notated as a1 and a3, one room, notated as a2, and one hall, notated as a4, are selected as target areas. Five APs (D-Link DAP 2310), notated as AP1, AP2, AP3, AP4, and AP5, and three MPs (SAMSUNG GT-S7568), notated as MP1, MP2, and MP3, are deployed in target areas for the testing. At the same time, the self-developed RSS signal collection software is used to collect the RSS data from all the hearable APs at each MP in the silence and intrusion states with the sampling rate of 1 Hz. The layout of the environment and the interface of RSS signal collection software are shown in Figure 4.

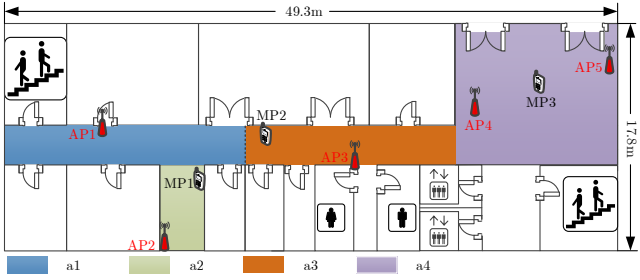

Fig. 3. Environmental layout.

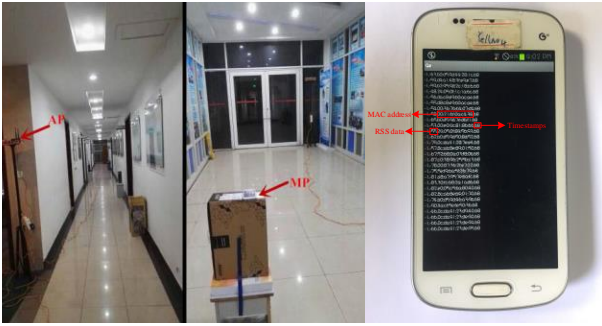

(a) Photos of the environment.   (b) Interface of signal collection software.
Fig. 4. Photos during the testing.

### B. Results analysis

In our testing, the following three indicators are adopted to evaluate the performance of WLAN indoor intrusion detection, i.e., False Positive (FP), the probability of judging as intrusion state in the absence of intrusion [37], False Negative (FN), the probability of judging as silence state in the presence of intrusion [38], and Detection Accuracy (DA), the probability of correct states judgment [39].

Besides, the following five kernel functions are chosen to build the kernel matrix to verify the advantages of the proposed transfer learning method, i.e., linear kernel $f_1(\mathbf{x}_i, \mathbf{x}_j) = \mathbf{x}_i^T \mathbf{x}_j$, Gaussian kernel $f_2(\mathbf{x}_i, \mathbf{x}_j) = \exp(-\gamma \|\mathbf{x}_i - \mathbf{x}_j\|^2)$, Laplace kernel $f_3(\mathbf{x}_i, \mathbf{x}_j) = \exp(-\sqrt{\gamma} \|\mathbf{x}_i - \mathbf{x}_j\|)$, inverse square distance kernel $f_4(\mathbf{x}_i, \mathbf{x}_j) = 1/(\gamma \|\mathbf{x}_i - \mathbf{x}_j\|^2 + 1)$, and inverse distance kernel $f_5(\mathbf{x}_i, \mathbf{x}_j) = 1/(\gamma \|\mathbf{x}_i - \mathbf{x}_j\| + 1)$, where $\gamma$ is the median distance between two different RSS fusion features in the source domain.

#### 1) Parameters discussion

In order to analyze the impact of the tradeoff parameters $\lambda$ and dimensions of transferred RSS features $q$ on the performance of the proposed method, Figure 5 shows the detection performance of the proposed method when the value $\lambda$ increases from 0.0001 to 1000 and the value $q$ increases from 20 to 100 under the condition of the slide window length $L = 20$. As can be seen from this figure, when the value $\lambda$ is small (e.g., $\lambda < 0.001$), that is, the complexity of the optimization transfer matrix $\mathbf{w}$ is relatively large, the performance of the iterative transfer is greatly affected by the change of RSS features, which leads to the reduction of the generalization ability of the system. When the value $\lambda$ is large (e.g., $\lambda > 0.1$), that is, the complexity of W is relatively small, it is difficult for the transfer matrix to transfer the RSS fusion features into the RKHS accurately, which affects the detection performance of the system. When $0.001 \leq \lambda \leq 0.1$, the proposed method can control the complexity of $\mathbf{w}$ and achieve better intrusion detection performance. Besides, increasing the value $q$ can expand the dimensions of the transfer matrix, which enables more preservation of RSS features. Sufficient samples of RSS features can be utilized to train classifiers to obtain more accurate classification result, which improves the stability of the proposed detection method. However, this process will increase the computational overhead of the system to a certain extent.

|  | Silence | a1 intrusion | a2 intrusion | a3 intrusion | a4 intrusion |  | Silence | a1 intrusion | a2 intrusion | a3 intrusion | a4 intrusion |
|---|---|---|---|---|---|---|---|---|---|---|---|
| Silence | 1.00 | 0.00 | 0.00 | 0.00 | 0.00 | Silence | 0.64 | 0.22 | 0.13 | 0.00 | 0.01 |
| a1 intrusion | 0.00 | 0.97 | 0.00 | 0.00 | 0.03 | a1 intrusion | 0.00 | 0.89 | 0.11 | 0.00 | 0.00 |
| a2 intrusion | 0.00 | 0.00 | 0.99 | 0.00 | 0.01 | a2 intrusion | 0.00 | 0.04 | 0.55 | 0.00 | 0.40 |
| a3 intrusion | 0.00 | 0.00 | 0.00 | 1.00 | 0.00 | a3 intrusion | 0.00 | 0.00 | 0.00 | 1.00 | 0.00 |
| a4 intrusion | 0.00 | 0.00 | 0.00 | 0.00 | 1.00 | a4 intrusion | 0.00 | 0.00 | 0.00 | 0.00 | 1.00 |

(a) Proposed Method   (b) Method without feature fusion
Fig. 7. Confusion matrices by different methods.

Figure 6 shows the confusion matrix of the system under different value $L$ when $\lambda = 0.1$ and $q = 40$ [40], in which the element on the $i$-th row and $j$-th column represents the probability that the $i$-th real state is judged as the $j$-th state. Obviously, with the increase of value $L$, the diagonal elements of this confusion matrix increase, which indicates the improvement of the detection performance.

In order to further illustrate the impact of the value $L$ on the detection performance, Figure 7 compares the confusion matrices of the proposed intrusion detection method and the method in which the RSS features are not optimized by the multi-branch deep convolutional neural network. Since the proposed method processes the original data and fuses the RSS features, the extracted RSS fusion features indicate the intrusion state in the target environment more effectively, and thereby it performs better in the complex indoor environment.

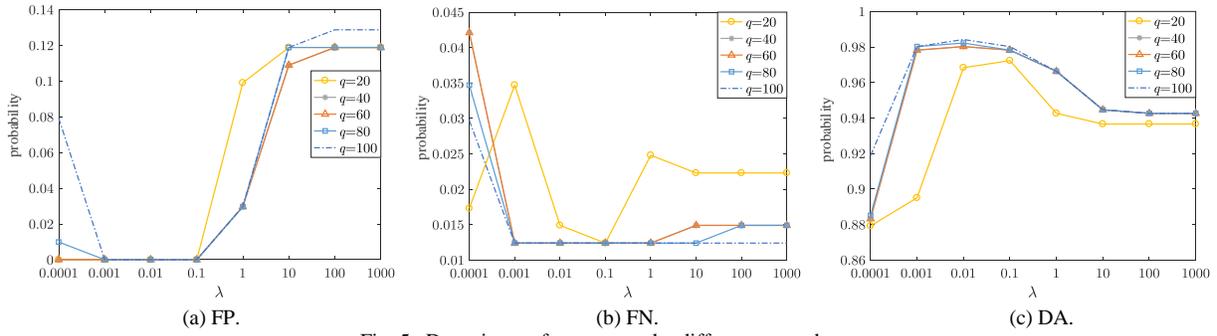

Fig. 5. Detection performance under different $\lambda$ and $q$.

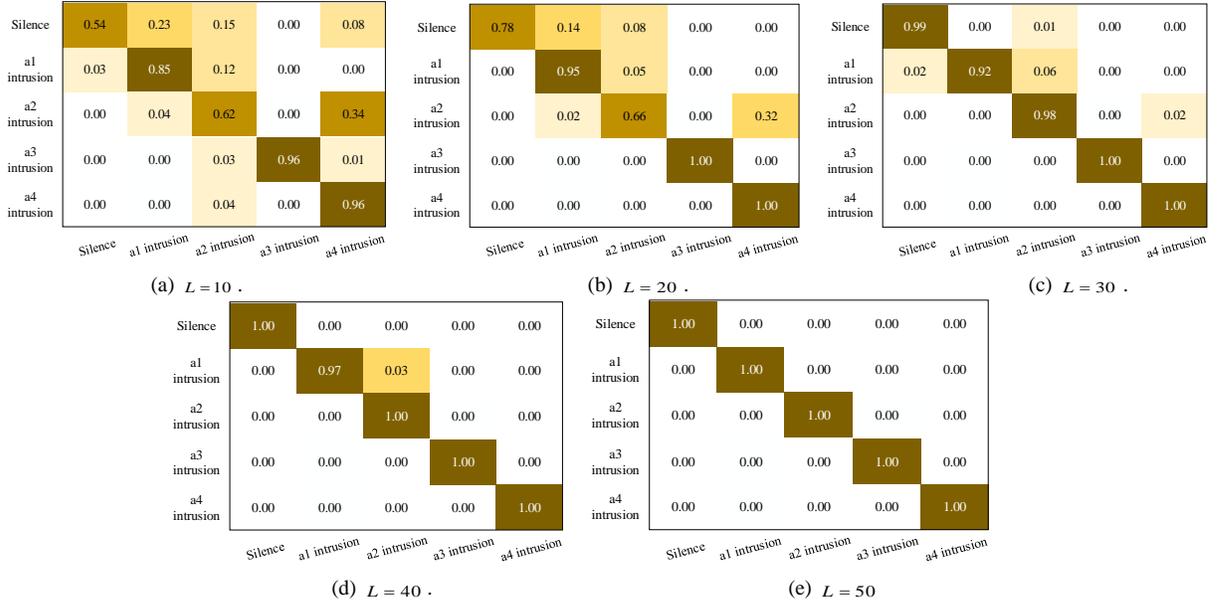

Fig. 6. Confusion matrix with different value $L$.

With the increase of value $L$, Figures 8 and 9 show the changes of FP, FN, and DA by the proposed method and the method changes of FP, FN, and DA by the proposed method and the method without RSS features optimization. On one hand, if the value $L$ is too small, the extracted RSS features cannot accurately depict the impact of intrusion targets on the signal fluctuation, which results in a decline in the detection performance. On the other hand, if the value $L$ is too large, the ability of extracted RSS features in perceiving silence and intrusion states in target areas is improved, but a large system delay is resulted. On the whole, the multi-branch deep convolutional neural network can be utilized to decrease the dimensions of collected RSS features, which can reduce the redundancy of signal features, and thereby the proposed method is more sensitive to the environmental change.

Figure 10 shows the changes of FP, FN, and DA by the proposed method with the increase of value $N$. Without the loss of generality, three classifiers, i.e., K-nearest Neighbor (KNN), Random Forest (RF), and Support Vector Machine (SVM), are trained by exploiting transferred RSS fusion features for intrusion detection, from which we can find that with the increase of value $N$, the detection performance of classifiers tends to converge on the whole.

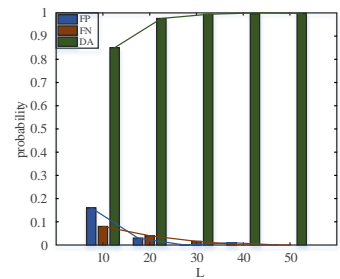

Fig. 8. Detection performance of the proposed method with different $L$.

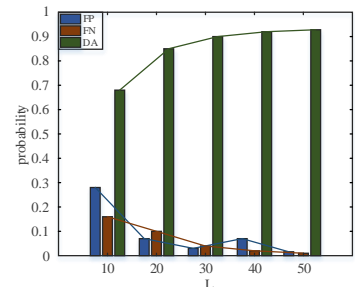

Fig. 9. Detection performance of the method without RSS features optimization with different $L$.

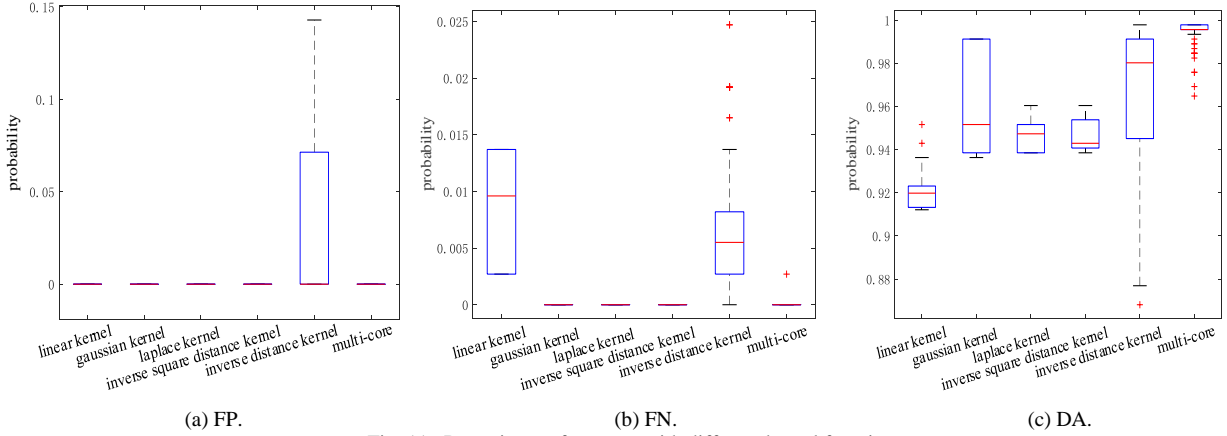

Fig. 11. Detection performance with different kernel functions.

Figure 11 shows the intrusion detection performance by the proposed method when choosing different kernel functions to construct the kernel matrix [41]. As can be seen from this figure, compared with the use of a single kernel function, the multi-core method can map different eigenvectors of RSS data through the most appropriate kernel function into the same shared subspace, where the RSS fusion features can be expressed more accurate and reasonable, which improves the classification accuracy of test samples and increases the robustness of the proposed method.

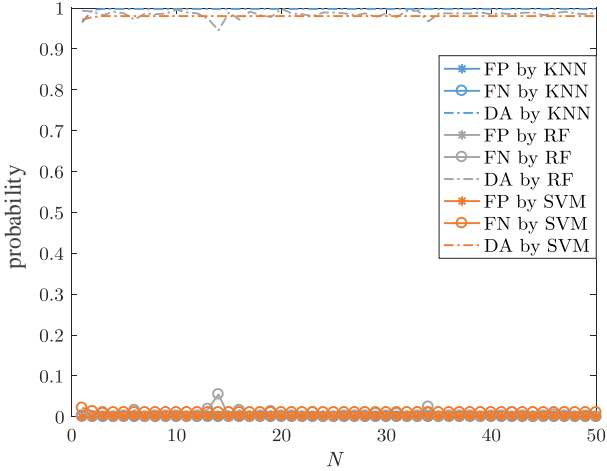

Fig. 10. Detection performance with different value $N$.

*2) Parameters discussion*

In Table 1, FP, FN, and DA are compared when classifiers are trained by the transferred and unmigrated RSS fusion features respectively. It can be seen that the detection accuracy of KNN, RF, and SVM trained by transferred RSS fusion features can reach 99.65%, 98.45%, and 98.13% respectively. On the whole, the system designed in this paper can capture the signal changes more sensitively in the indoor environment and shows higher stability.

Besides, the intrusion detection performance by the proposed method is compared with the existing RASID, PNN, and PRNN methods in Table 2. It can be seen that the proposed method has the best detection performance on the whole, and its DA is increased by 6.32%, 5.38%, and 4.18% respectively compared with the other three methods. Therefore, it is verified that the proposed method can improve the detection performance of the system by reducing the overall difference between the RSS features of source and target domains.

TABLE I
DETECTION PERFORMANCE WITH DIFFERENT CLASSIFIERS

| categories | FP | FN | DA |
|---|---|---|---|
| KNN (before transfer) | 34.87% | 0 | 76.11% |
| KNN (after transfer) | 0 | 0 | 99.65% |
| RF (before transfer) | 6.53% | 1.62% | 83.96% |
| RF (after transfer) | 0 | 0 | 98.45% |
| SVM (before transfer) | 17.99% | 0 | 93.85% |
| SVM (after transfer) | 0 | 1.10% | 98.13% |

TABLE II
DETECTION PERFORMANCE BY DIFFERENT METHODS

| Metrics | RASID | PNN | PRNN | Proposed |
|---|---|---|---|---|
| FP | 6.81% | 3.39% | 0 | 0 |
| FN | 3.27% | 2.77% | 0 | 0 |
| DA | 93.41% | 94.28% | 95.53% | 99.65% |

## V. CONCLUSION

By considering that WLAN signals are usually time-varying and contain noise in the complex indoor environment, this paper proposes a new WLAN indoor intrusion detection method based on the deep signal feature fusion and minimized-MKMMD transfer learning. Firstly, the multi-branch deep convolutional neural network is utilized to conduct feature fusion for RSS data, and then the optimal transfer matrix based on the MKMMD minimization is constructed. Secondly, the optimal migration matrix is used to transfer the RSS fusion features of source and target domains into the same shared subspace to reduce their difference in the same location. Finally, the intrusion detection in the target environment is realized by iteratively updating the process above. Experimental results show that the proposed method can eliminate the interference of the WLAN signal volatility on RSS features to some extent, which effectively enhances the robustness of the classifiers used for intrusion detection. However, how to use the proposed method to detect the static and dynamic objects simultaneously forms an interesting work in future.

## REFERENCES

[1] S. Vishwakarma and A. Agrawal, "A survey on activity recognition and behavior understanding in video surveillance," *Visual Computer*, vol. 13, no. 1, pp. 983-1009, 2013.

[2] S. Biswas and P. Milanfar, "Linear support tensor machine with LSK channels: Pedestrian detection in thermal infrared images," *IEEE Transactions on Image Processing*, vol. 26, no. 9, pp. 4229-4242, 2017.

[3] E. Felemban, F. Shaikh, U. Qureshi, and et al, "Underwater sensor network applications: A comprehensive survey," *International Journal of Distributed Sensor Networks*, vol. 2015, no. 5, pp. 1-14, 2015.

[4] Z. Chen, M. Li, and Y. Tan, "Perception-aware multiple scalable video streaming over WLANs," *IEEE Signal Processing Letters*, vol. 17, no. 7, pp. 675-678, 2010.

[5] F. Zhang, Q. Shen, X. Shi, and et al, "Infrared detection based on localized modification of Morpho butterfly wings," *Advanced Materials*, vol. 27, no. 6, pp. 1077-1082, 2015.

[6] J. Wang, Z. Li, M. Li, and et al, "Sensor network navigation without locations," *IEEE Transactions on Parallel and Distributed Systems*, vol. 24, no. 7, pp. 1436-1446, 2013.

[7] O. Said, and A. Elnashar, "Scaling of wireless sensor network intrusion detection probability: 3D sensors, 3D intruders, and 3D environments," *EURASIP Journal on Wireless Communications and Networking*, vol. 2015, no. 1, pp. 1-12, 2015.

[8] M. Zhou, Q. Zhang, Z. Tian, and et al, "Simultaneous pathway mapping and behavior understanding with crowdsourced sensing in WLAN environment," *Ad Hoc Networks*, vol. 58, no. 4, pp. 160-170, 2016.

[9] S. Choi S, K. Park, and C. Kim, "Performance impact of interlayer dependence in infrastructure WLANs," *IEEE Transactions on Mobile Computing*, vol. 5, no. 7, pp. 829-845, 2006.

[10] R. Bace. "Intrusion detection," *Computerworld*, vol. 3, no. 1, pp. 1-29, 1999.

[11] J. Lv, D. Man, W. Yang, and et al, "Robust WLAN-based indoor intrusion detection using PHY layer information," *IEEE Access*, vol. 7, pp. 30117-30127, 2017.

[12] A. Kosba, A. Saeed, and M. Youssef, "RASID: A robust WLAN device-free passive motion detection system," *IEEE International Conference on Pervasive Computing and Communications*, pp. 180-189, 2012.

[13] Z. Tian, X. Zhou, M. Zhou, and et al, "Indoor device-free passive localization for intrusion detection using multi-feature PNN," *International Conference on Communications and Networking*, pp. 272-277, 2015.

[14] L. Li, H. Zhang, H. Peng, and et al, "Nearest neighbors based density peaks approach to intrusion detection," *Chaos Solitons and Fractals*, vol. 110, no. 5, 2018.

[15] R. Vinayakumar, M. Alazab, K. Soman, and et al, "Deep learning approach for intelligent intrusion detection system," *IEEE Access*, vol. 7, pp. 41525-41550, 2019.

[16] S. Musavi, and M. Hashemi. "HPCgnature: a hardware-based application-level intrusion detection system," *IET Information Security*, vol. 13, no. 1, pp. 19-26, 2019.

[17] Q. Tan, C. Han, L. Sun L, and et al, "A CSI frequency domain fingerprint-based method for passive indoor human detection," *IEEE International Conference on Trust, Security and Privacy in Computing and Communications*, pp. 1832-1837, 2018.

[18] Y. Li, L. Wu, P. Wouters, and et al, "Effect of ground return path on partial discharge signal propagation along single‐core and three‐core power cables," *International Transactions on Electrical Energy Systems*, vol. 26, no. 8, 2016.

[19] S. Krishnan, "The Tau model for data eedundancy and information combination in earth sciences: Theory and application," *Mathematical Geosciences*, vol. 40, no. 6, pp. 705-727, 2008.

[20] G. Cheng, Y. Wang, S. Xu, and et al, "Automatic road detection and centerline extraction via cascaded end-to-end convolutional neural network," *IEEE Transactions on Geoscience and Remote Sensing*, vol. 55, no. 6, pp. 3322-3337, 2017.

[21] I. Caldwell, M. Correia, J. Palma, and et al, "Advances in tagging syngnathids, with the effects of dummy tags on behaviour of hippocampus guttulatus," *Journal of Fish Biology*, vol. 78, no. 6, pp. 1769-1785, 2011.

[22] N. Chaudhuri, B. Chaudhuri, S. Ray, and et al, "Wide-area phasor power oscillation damping controller: A new approach to handling time-varying signal latency," *Iet Generation Transmission and Distribution*, vol. 4, no. 5, pp. 620-630, 2010.

[23] T. Erven, and P. Harremos, "Divergence and Kullback-Leibler divergence." *IEEE Transactions on Information Theory*, vol. 60, no. 7, pp. 3797-3820, 2014.

[24] K. Borgwardt, A. Gretton, and M. Rasch, and et al, "Integrating structured biological data by Kernel Maximum Mean Discrepancy," *Bioinformatics*, vol. 22, no. 14, pp. e49-e57, 2006.

[25] O. Arqub, M. Al-Smadi, S. Momani, and et al, "Numerical solutions of fuzzy differential equations using reproducing kernel Hilbert space method," *Soft Computing*, vol. 20, no. 8, pp. 3282-3302, 2016.

[26] A. Gretton, B. Sriperumbudur, D. Sejdinovic, and et al, "Optimal kernel choice for large-scale two-sample tests," *International Conference on Neural Information Processing Systems*, pp. 1025-1213, 2012.

[27] P. Du, K. Tan, and X. Xing, "Wavelet SVM in Reproducing Kernel Hilbert Space for hyperspectral remote sensing image classification," *Optics Communications*, vol. 283, no. 24, pp. 4978-4984, 2010.

[28] A. Coluccia and F. Ricciato. "A Software-Defined Radio tool for experimenting with RSS measurements in IEEE 802.15.4: Implementation and applications," *International Journal of Sensor Networks*, vol. 14, no. 3, pp. 144-154, 2013.

[29] W. Bian and D. Tao. "Max-Min Distance analysis by using sequential SDP Relaxation for dimension reduction," *IEEE Transactions on Pattern Analysis and Machine Intelligence*, vol. 33, no. 5, pp. 1037-1050, 2011.

[30] J. Shawe-Taylor, C. Williams, N. Cristianini, and et al, "On the eigenspectrum of the gram matrix and the generalization error of kernel-PCA," I*EEE Transactions on Information Theory*, vol. 7, no. 7, pp. 2510-2522, 2005.

[31] B. hölkopf, A. Smola, and K. Müller, "Nonlinear Component analysis as a Kernel Eigenvalue problem," *Neural Computation*, vol. 10, no. 5, pp. 1299-1319, 1998

[32] M. Schoeberl, S. Abbaspour, and B. Akesson, and et al, "T-CREST: Time-predictable multi-core architecture for embedded systems," *Journal of Systems Architecture*, vol. 61, no. 9, pp. 449-471, 2015.

[33] Y. Chang, F. Nie, and M. Wang. "Multiview Feature analysis via structured sparsity and Shared Subspace discover," *Neural Computation*, vol. 29, no. 7, pp. 1986-2003, 2017.

[34] S. Silvey. "The Lagrangian Multiplier Test," *Annals of Mathematical Statistics*, vol. 30, no. 2, pp. 389-407, 1959.

[35] W. Zou, Y. Feng, J. Dong, and et al, "A new strategy to iteratively update scalable universal quantitative models for the testing of azithromycin by near infrared spectroscopy," *Science China Chemistry*, vol. 56, no. 4, pp. 533-540, 2013.

[36] P. Beerli, "Maximum likelihood estimation of a migration matrix and effective population sizes in n subpopulations by using a coalescent approach," *Proc Natl Acad Sci U S A*, vol. 98, no, 8, pp. 4563-4568, 2001.

[37] P. Bonoris, P. Greenberg, G. Christison, and et al, "Ability of R-wave amplitude (RWA) changes to reduce false negative (FN) and false positive (FP) responses by ST depression in treadmill stress testing (TST)," *American Journal of Cardiology*, vol. 41, no. 2, pp. 378-378, 1978.

[38] J. Ioannidis and J. Mclaughlin, "The false-positive to false-negative ratio in epidemiologic studies," *Epidemiology*, vol. 22, no. 4, pp. 450-456, 2011.

[39] L.Wu, Z. Ling, L. Jiang, and et al, "Long-Range surface plasmon with graphene for enhancing the sensitivity and detection accuracy of biosensor," *IEEE Photonics Journal*, vol. 8, no. 2, pp. 1-9, 2016.

[40] R. Salla, H. Wilhelmiina, K. Sari, and et al, "Evaluating dynamic reliability of sensors based on evidence theory and confusion matrix," *Control and Decision*, vol. 30, no. 6, pp. 1111-1115, 2015.

[41] M. Li, W. Bi, J. Kwok, and et al, "Large-scale Nystrom kernel matrix approximation using randomized SVD," *IEEE Transactions on Neural Networks and Learning Systems*, vol. 26, no. 1, pp. 152-164, 2014.

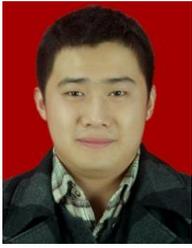

**Mu Zhou** received the Ph.D. degree in Communication and Information Systems from the Harbin Institute of Technology (HIT), China, in 2012. He was a Joint-cultivated Ph.D. Student at the University of Pittsburgh (PITT), USA, and a Post-Doctoral Research Fellow at the Hong Kong University of Science and Technology (HKUST), China. Afterward, he joined Chongqing University of Posts and Telecommunications (CQUPT), China, where he has been a Full Professor with the School of Communication and Information Engineering since 2014. He was supported by the Chongqing Municipal Program of Top-Notch Young Professionals for Special Support of Eminent Professionals and awarded with Outstanding Cooperation Project Award from Huawei Technologies Co., Ltd. Over the past 5 years, he was engaged in 5 national projects, 9 provincial and ministerial projects (including 2 major projects), and 7 enterprise projects. His main research areas include wireless localization and navigation, signal reconnaissance and detection, and convex optimization and deep learning. He has published more than 100 peer-review research papers and served on Technical Program Committees of IEEE ICC, GLOBECOM, WCNC, IWCMC, VTC, IWCMC, and et al. He is a Senior Member of IEEE.

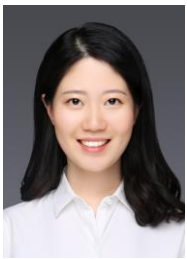

**Xinyue Li** received the B.S. degree in Telecommunication Engineering from the Chongqing University of Posts and Telecommunications, China, in 2016, where she is currently pursuing the M.S. degree. Her current research interests include indoor intrusion detection, transfer learning, and maximum mean discrepancy.

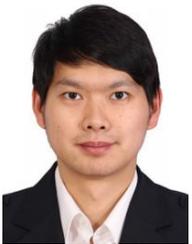

**Yong Wang** (M'18) received the B.S., M.S. and Ph.D. degrees from Harbin Institute of Technology, Harbin, China, in 2010, 2012 and 2018, respectively. He is currently a Lecturer with Chongqing University of Posts and Telecommunications, Chongqing, China. From January 2014 to June 2015, he was a visiting Ph.D. student at the Department of Electrical and Computer Engineering, University of Toronto, Canada. His research interests include resource allocation and signal processing in cooperative networks, deep learning and WIFI localization.

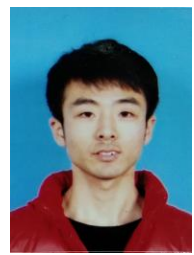

**Aihu Ren** received the B.S. degree in Telecommunication Engineering from the Shangqiu Normal University, China, in 2018, where he is currently pursuing the M.S. degree. His current research interests include indoor localization, machine learning, and deep neural network.

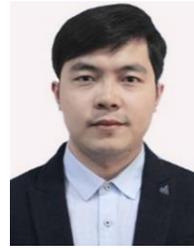

**Xiaolong Yang** received his M.Sc. degree and Ph.D. degree in communication engineering from Harbin Institute of Technology in 2012 and 2017 respectively. From 2015 to 2016, he was a visiting scholar at Nanyang Technological University, Singapore. He is currently a lecturer in Chongqing University of Posts and Telecommunications. His current research interests include cognitive radio networks and energy efficiency optimization.